\documentstyle[12pt,axodraw]{article}
\begin{document}
\baselineskip 0.6cm

\def\simgt{\mathrel{\lower2.5pt\vbox{\lineskip=0pt\baselineskip=0pt
           \hbox{$>$}\hbox{$\sim$}}}}
\def\simlt{\mathrel{\lower2.5pt\vbox{\lineskip=0pt\baselineskip=0pt
           \hbox{$<$}\hbox{$\sim$}}}}
\newcommand{\vev}[1]{ \langle {#1} \rangle }

\begin{titlepage}

\begin{flushright}
UCB-PTH-06/06 \\
LBNL-60061
\end{flushright}

\vskip 1.7cm

\begin{center}

{\Large \bf 
Holographic Grand Unification
}

\vskip 1.0cm

{\large
Yasunori Nomura, David Poland and Brock Tweedie
}

\vskip 0.4cm

{\it Department of Physics, University of California,
           Berkeley, CA 94720} \\
{\it Theoretical Physics Group, Lawrence Berkeley National Laboratory,
           Berkeley, CA 94720} \\

\vskip 1.2cm

\abstract{We present a framework for grand unification in which 
the grand unified symmetry is broken spontaneously by strong gauge 
dynamics, and yet the physics at the unification scale is described 
by (weakly coupled) effective field theory.  These theories are 
formulated, through the gauge/gravity correspondence, in truncated 
5D warped spacetime with the UV and IR branes setting the Planck and 
unification scales, respectively.  In most of these theories, the 
Higgs doublets arise as composite states of strong gauge dynamics, 
corresponding to degrees of freedom localized to the IR brane, 
and the observed hierarchies of quark and lepton masses and mixings 
are explained by the wavefunction profiles of these fields in the 
extra dimension.  We present several realistic models in this framework. 
We focus on one in which the doublet-triplet splitting of the Higgs 
fields is realized within the dynamical sector by the pseudo-Goldstone 
mechanism, with the associated global symmetry corresponding to 
a bulk gauge symmetry in the 5D theory.  Alternatively, the light 
Higgs doublets can arise as a result of dynamics on the IR brane, 
without being accompanied by their triplet partners.  Gauge coupling 
unification and proton decay can be studied in these models using 
higher dimensional effective field theory.  The framework also sets 
a stage for further studies of, e.g., proton decay, fermion masses, 
and supersymmetry breaking.}

\end{center}
\end{titlepage}

\section{Introduction}
\label{sec:intro}

Weak scale supersymmetry (SUSY) provides an elegant solution to the 
naturalness problem of the standard model, by invoking a cancellation 
between the standard model and its superpartner contributions to the 
Higgs potential.  An interesting consequence of this framework is that 
the three gauge couplings unify at an extremely high energy of order 
$M_U \approx 10^{16}~{\rm GeV}$, if a normalization of the $U(1)_Y$ 
gauge coupling is adopted that allows the embedding of the standard 
model gauge group into a larger simple symmetry group: $SU(5) \supset 
SU(3)_C \times SU(2)_L \times U(1)_Y$.  This suggests the existence 
of some unified physics above this energy scale, which in some form 
utilizes $SU(5)$ or a larger group containing it.

The simplest possibilities for physics above $M_U$ are four dimensional 
(4D) supersymmetric grand unified theories (GUTs)~\cite{Dimopoulos:1981zb}. 
In these theories, physics above $M_U$ is described by 4D supersymmetric 
gauge theories in which the standard model gauge group is embedded 
into a larger (simple) gauge group.  This, however, leads to the 
problem of doublet-triplet splitting in the Higgs sector, and often 
leads to too rapid proton decay caused by the exchange of colored 
triplet Higgsinos~\cite{Murayama:2001ur}.  While several solutions 
to these problems have been proposed within conventional 4D SUSY 
GUTs~[\ref{Witten:1981kv:X}~--~\ref{Yanagida:1994vq:X}], their 
explicit implementations often require the introduction of a larger 
multiplet(s) and/or specifically chosen superpotential interactions, 
especially when one tries to make the models fully realistic.  This 
loses a certain beauty the simplest theory had, especially if one 
adopts the viewpoint that these theories are ``fundamental,'' arising 
directly from physics at the gravitational scale, e.g. string theory.

An alternative possibility for physics above $M_U$ is that the unified 
gauge symmetry is realized in higher dimensional (semi-)classical 
spacetime~[\ref{Kawamura:2000ev:X}~--~\ref{Hall:2002ea:X}].  In this 
case there is no 4D unified gauge symmetry containing the standard 
model gauge group as a subgroup --- the unified symmetry in higher 
dimensions is broken locally and explicitly by a symmetry breaking 
defect.  This structure allows a natural splitting between the doublet 
and triplet components for the Higgs fields, while the successful 
prediction for gauge coupling unification is recovered by diluting 
the effects from the defect due to a moderately large extra dimension(s). 
Dangerous proton decay is suppressed by an $R$ symmetry, arising 
naturally from the higher dimensional structure of the triplet Higgsino 
mass matrix.  The framework also allows for a simple understanding 
of the observed structure of fermion masses and mixings, in terms 
of wavefunction suppressions of the Yukawa couplings arising for 
bulk quarks and leptons~\cite{Hall:2001zb,Hebecker:2002re}.

In this paper we study a framework for physics above $M_U$ in 
which the standard model gauge group is unified into a simple gauge 
group in precisely the same sense as in conventional 4D SUSY GUTs, 
and yet mechanisms and intuitions developed in higher dimensions 
can be used to address the various issues of unified theories. 
Let us consider that the standard model gauge group is embedded into 
a simple unified gauge group, e.g. $SU(5)$, at energies above $M_U$. 
We assume that the unified gauge symmetry is broken by strong gauge 
dynamics associated with another gauge group $G$, and that this gauge 
group has a large 't~Hooft coupling $\tilde{g}^2 \tilde{N}/16\pi^2 
\gg 1$, where $\tilde{g}$ and $\tilde{N}$ are the gauge coupling 
and the number of ``colors'' for the gauge group $G$.  With these 
values of the 't~Hooft coupling for $G$, an appropriate (weakly 
coupled) description of physics is given in higher dimensional 
warped spacetime (for $\tilde{N} \gg 1$), due to the gauge/gravity 
correspondence~\cite{Maldacena:1997re}. In the simplest setup where 
$\tilde{g}$ evolves slowly above the dynamical scale, our theories are 
formulated in 5D anti-de~Sitter (AdS) spacetime truncated by two branes, 
where the curvature scales on the ultraviolet (UV) and infrared (IR) 
branes are chosen to be $k \approx (10^{17}-10^{18})~{\rm GeV}$ and 
$k' \approx (10^{16}-10^{17})~{\rm GeV}$, respectively.  This allows 
us to construct simple ``calculable'' unified theories in which the 
unified gauge symmetry is broken dynamically --- physics above $M_U$ 
is determined simply by specifying parameters in higher dimensional 
effective field theory.

In this paper we construct realistic unified theories in the 
framework described above.  In general, there are many ways to address 
the issues of unified theories in our framework.  In one example, 
which we discuss in detail, we use the idea that the Higgs doublets 
of the minimal supersymmetric standard model (MSSM) are pseudo-Goldstone 
bosons of a broken global symmetry~\cite{Inoue:1985cw,Barbieri:1992yy}. 
Specifically, we assume that the $G$ sector possesses a global $SU(6)$ 
symmetry, of which an $SU(5)$ ($\times U(1)$) subgroup is gauged. 
The gauged $SU(5)$ group contains the standard model gauge group as 
a subgroup.  We assume that the dynamics of $G$ breaks the global 
$SU(6)$ symmetry down to $SU(4) \times SU(2) \times U(1)$ at the scale 
$M_U$, which leads to the correct gauge symmetry breaking, $SU(5) 
\rightarrow SU(3)_C \times SU(2)_L \times U(1)_Y$, and ensures that 
the Higgs doublets remain massless after the symmetry breaking, without 
being accompanied by their colored triplet partners.  The simplest 
realization of our theory, corresponding to this symmetry structure, 
is then obtained in 5D truncated warped space in which the bulk $SU(6)$ 
gauge symmetry is broken to $SU(5) \times U(1)$ and $SU(4) \times SU(2) 
\times U(1)$ on the UV and IR branes, respectively.  Realistic unified 
theories having this symmetry structure were constructed previously 
in flat space in Ref.~\cite{Burdman:2002se}, where the symmetry 
breakings on the two branes are both caused by boundary conditions, 
and in Ref.~\cite{Cheng:1999fw}, where the breakings are by the 
Higgs mechanism.  In our context, we find that the simplest theory 
is obtained if the breakings on the UV and IR branes are caused by 
boundary conditions and the Higgs mechanism, respectively.  Note that, 
in the ``4D description'' of the theory, the Higgs breaking on the IR 
brane corresponds to dynamical GUT breaking, and the low-energy Higgs 
doublets are interpreted as composite particles of the dynamical 
GUT-breaking sector.  This theory thus provides a simple explicit 
realization of the composite pseudo-Goldstone Higgs doublets, in which 
the origin of the global $SU(6)$ symmetry can be understood as the 
``flavor'' symmetry of the dynamical GUT-breaking sector. 

Below the GUT-breaking scale $M_U$, our theory is reduced to the 
MSSM (supplemented by small seesaw neutrino masses).  The successful 
unification prediction for the low-energy gauge couplings is 
preserved as long as the threshold corrections from the dynamical 
GUT-breaking sector are sufficiently small.  Our higher dimensional 
description of the theory allows us to estimate the size of these 
corrections, and we find that this can be the case.  Dimension five 
proton decay does not exclude the theory, because of the existence 
of these threshold corrections.  Realistic quark and lepton mass 
matrices can also be reproduced, where the observed hierarchies 
in masses and mixings are understood in terms of the wavefunction 
profiles of the quark and lepton fields.  In the 4D description of 
the theory, these hierarchies arise through mixings between elementary 
states and composite states of $G$, which are given by powers of 
$M_U/M_*$, where $M_*$ is the fundamental scale of the theory, close 
to the 4D Planck scale.  Unwanted unified mass relations for the 
first two generation fermions do not arise, because of GUT breaking 
effects in the $G$ sector.

We also discuss other possible theories in our framework.  We show 
that it allows for the construction of large classes of models, 
including missing partner type and product group type models. 
In most of them, the Higgs doublets arise as states localized to 
the IR brane, corresponding to composite states of the strong $G$ 
dynamics. A 4D scenario related to these theories was discussed 
previously in Ref.~\cite{Kitano:2005ez}, based on a supersymmetric 
conformal field theory (CFT), where a possible AdS interpretation 
was also noted.  In all of these theories, our higher dimensional 
framework allows a straightforward implementation of the mechanism 
generating the hierarchical fermion masses and mixings, in terms of 
the wavefunction profiles of matter fields in the extra dimension.

The organization of the paper is as follows.  In the next section we 
describe the basic structure of our theory using the 4D description. 
We describe how the MSSM arises naturally at low energies in this 
theory.  In section~\ref{sec:model} we construct an explicit model 
in truncated 5D warped space.  We show that the model does not suffer 
from problems of conventional 4D SUSY GUTs, e.g. the doublet-triplet 
splitting and dimension five proton decay problems, and also that 
the observed hierarchies in the quark and lepton mass matrices can 
be understood in terms of the wavefunction profiles of these fields 
in the extra dimension.  In section~\ref{sec:other}, we discuss 
other possible theories in our framework, including missing partner 
type and product group type models.  Discussion and conclusions 
are given in section~\ref{sec:concl}, which include a comment 
on the possibility of having a theory with $\tilde{g}^2 
\tilde{N}/16\pi^2 \simlt 1$.

\section{Basic Picture}
\label{sec:picture}

In this section we describe our theory using the 4D description. 
Here we focus on the case where the light Higgs doublets of the MSSM 
arise as pseudo-Goldstone supermultiplets of the GUT scale dynamics. 
This has the virtue that the success of the theory is essentially 
guaranteed by its symmetry structure, without relying on specifically 
chosen matter content or interactions.  Other possibilities will be 
discussed in section~\ref{sec:other}.

We consider that the standard model gauge group is embedded into 
a simple gauge group $SU(5)$, which is spontaneously broken at 
the scale $M_U \approx 10^{16}~{\rm GeV}$.  What is the underlying 
dynamics of this symmetry breaking?  A hint will come from considering 
how the MSSM arises below the symmetry breaking scale $M_U$.  In 
particular, considering how the MSSM matter content naturally appears 
at energies below $M_U$ and why interactions among these particles 
-- the gauge and Yukawa interactions -- take the observed form and 
values will provide a guide to the physics of this symmetry breaking. 
The suppression of certain operators allowed by standard model gauge 
invariance, e.g. the ones leading to dangerous dimension five proton 
decay, may also give hints regarding the structure of this physics.

We focus on the possibility that the unified gauge group, $SU(5)$, is 
spontaneously broken by dynamics associated with another gauge group 
$G$.  In this setup, the $G$ sector is charged under $SU(5)$, as it 
breaks $SU(5)$ dynamically.  The setup also allows the existence of other 
fields -- elementary fields -- that are singlet under $G$ and charged 
under $SU(5)$.  Suppose now that the theory has a matter content that 
satisfies $n_{{\bf 5}^*} - n_{\bf 5} = n_{\bf 10} - n_{{\bf 10}^*} = 3$ 
and $n_{\bf r} - n_{{\bf r}^*} = 0$ (${\bf r} \neq {\bf 5}, {\bf 10}$), 
where $n_{\bf r}$ represents the number of $SU(5)$ multiplets in 
a complex representation ${\bf r}$.  The matter content is arbitrary 
otherwise.  (Note that this is not a very strong requirement on the 
spectrum --- with $n_{\bf r} - n_{{\bf r}^*} = 0$ for ${\bf r} \neq 
{\bf 5}, {\bf 10}$, the condition $n_{{\bf 5}^*} - n_{\bf 5} = n_{\bf 10} 
- n_{{\bf 10}^*}$ arises automatically as a consequence of anomaly 
cancellation.)  With this assumption, the low energy matter content 
is expected to be just the three generations of quarks and leptons, 
no matter what happens associated with the dynamics of the GUT-breaking 
sector $G$.  In general, the gauge dynamics of $G$ will produce an 
arbitrary number of split GUT multiplets as composite states, by picking 
up the effect of GUT breaking.  These states can then mix with the 
elementary states, so that the low energy states are in general mixtures 
of elementary and composite states and thus a collection of various 
incomplete $SU(5)$ multiplets.  Nevertheless, conservation of chirality 
guarantees that we always have three generations of quarks and leptons 
at low energies, although they may not arise simply from three copies 
of $({\bf 5}^* + {\bf 10})$.  Assuming that all the fields vector-like 
under the standard model gauge group obtain masses of order $M_U$ 
through nonperturbative effects of $G$, the matter content below 
$M_U$ is exactly the three generations of quarks and leptons.

The above argument shows that we can naturally obtain a low-energy 
chiral matter content that fills complete $SU(5)$ multiplets for chirality 
reasons (although each component in a multiplet may come from several 
different $SU(5)$ multiplets at high energies).  It also implies that 
any multiplets that do not fill out a complete $SU(5)$ multiplet must 
be vector-like.  It is interesting that the MSSM has exactly this 
structure.  Unless there is some special reason, however, the vector-like 
states are all expected to have masses of order $M_U$ from nonperturbative 
effects of $G$.  What could the special reason be for the Higgs doublets? 

The lightness of the Higgs doublets can be understood group theoretically 
if we identify these states as pseudo-Goldstone bosons of a broken global 
symmetry~\cite{Inoue:1985cw}.  Suppose that the $G$ sector possesses 
a global $SU(6)$ symmetry, of which an $SU(5)$ ($\times U(1)$) subgroup 
is gauged and identified as the unified gauge symmetry.  We assume that 
the dynamics of $G$ breaks the global $SU(6)$ symmetry down to $SU(4) 
\times SU(2) \times U(1)$ at the dynamical scale $\approx M_U$ in such 
a way that the gauged $SU(5)$ subgroup is broken to the standard model 
gauge group $SU(3)_C \times SU(2)_L \times U(1)_Y$ (321).  This leads 
to Goldstone chiral supermultiplets, whose quantum numbers under 321 
are given by $({\bf 3}, {\bf 2})_{-5/6} + ({\bf 3}^*, {\bf 2})_{5/6} 
+ ({\bf 1}, {\bf 2})_{1/2} + ({\bf 1}, {\bf 2})_{-1/2}$.  While the 
first two of these are absorbed by the broken $SU(5)$ gauge multiplets 
(the massive XY gauge supermultiplets), the last two are left in the 
low energy spectrum.  Although the global $SU(6)$ symmetry of the $G$ 
sector is explicitly broken by the gauging of the $SU(5)$ ($\times U(1)$) 
subgroup, the supersymmetric nonrenormalization theorem guarantees that 
the mass term for $({\bf 1}, {\bf 2})_{1/2} + ({\bf 1}, {\bf 2})_{-1/2}$ 
is not generated without picking up the effect of supersymmetry breaking, 
allowing us to identify these states as the two Higgs doublets of the 
MSSM: $H_u({\bf 1}, {\bf 2})_{1/2}$ and $H_d({\bf 1}, {\bf 2})_{-1/2}$. 
This provides a complete understanding of the MSSM field content in 
our framework.  The MSSM states -- the gauge, matter and Higgs fields 
-- are the only states that could not get a mass of order $M_U$ from 
$G$, because they are protected by gauge invariance, chirality, and 
the (pseudo-)Goldstone mechanism.

Since the two Higgs doublets arise from the dynamical breaking of 
$SU(6)$, they are composite states of $G$.  Suppose now that the 
dynamics of $G$ also produces composite states that have the same 321 
quantum numbers as the MSSM quarks and leptons, ${\cal Q}, {\cal U}, 
{\cal D}, {\cal L}$ and ${\cal E}$.  These composite states will then 
have ``Yukawa couplings'' with the Higgs fields at $M_U$, $W \approx 
{\cal Q} {\cal U} H_u + {\cal Q} {\cal D} H_d + {\cal L} {\cal E} H_d$, 
where the sizes of the couplings are naturally of order $4\pi$.  These 
couplings, however, disappear at low energies after integrating out all 
the heavy modes, because the strong $G$ dynamics respects $SU(6)$ and 
the Higgs doublets are the Goldstone bosons associated with the dynamical 
breaking of $SU(6)$.  Now, suppose that the theory also has several 
elementary fields that transform as ${\bf 5}^*$ and ${\bf 10}$ under 
$SU(5)$.  In this case the low-energy quarks and leptons, 
$Q, U, D, L$ and $E$, are in general linear combinations of the 
elementary and composite states.  The Yukawa couplings for these 
low-energy fields, $W \approx Q U H_u + Q D H_d + L E H_d$, can then 
be nonzero because the elementary states do not respect the full 
$SU(6)$ symmetry.  The sizes of the Yukawa couplings are determined by 
the strengths of the mixings between the elementary and composite states, 
which are in turn determined by the dimensions of the $G$-invariant 
operators that interpolate the composite states.  This situation is 
analogous to the case where the standard model Higgs boson is identified 
as a pseudo-Goldstone boson of strong gauge dynamics at the TeV 
scale~\cite{Contino:2003ve}.  By choosing operator dimensions to be 
larger for lighter generations, we can naturally understand the origin 
of the hierarchical structure for the quark and lepton masses and 
mixings.  The unwanted mass relations for the quarks and leptons 
can be avoided because low-energy quarks and leptons feel the 
GUT-breaking effects in the $G$ sector.

Dangerous dimension four and five proton decay can be suppressed if 
the theory possesses a continuous or discrete $R$ symmetry, under which 
the low-energy MSSM fields carry the charges $Q(1)$, $U(1)$, $D(1)$, 
$L(1)$, $E(1)$, $H_u(0)$ and $H_d(0)$ (and $N(1)$ if we introduce 
right-handed neutrino superfields $N$).  This $R$ symmetry is most 
likely spontaneously broken by the dynamics of the $G$ sector (unless 
there is a low-energy singlet field that transforms nonlinearly 
under this symmetry; see discussion in section~\ref{sec:other}). 
The $R$ symmetry should also be broken to the $Z_2$ subgroup, 
the $R$ parity of the MSSM, in order to give weak scale masses to 
the gauginos.  Supersymmetry breaking produces supersymmetric and 
supersymmetry-breaking masses for the Higgs doublets, as well as 
masses for the gauginos, squarks and sleptons, ensuring the stability 
of the desired vacuum.  Successful supersymmetric gauge coupling 
unification is preserved if the threshold corrections associated 
with the $G$ sector are sufficiently small. 

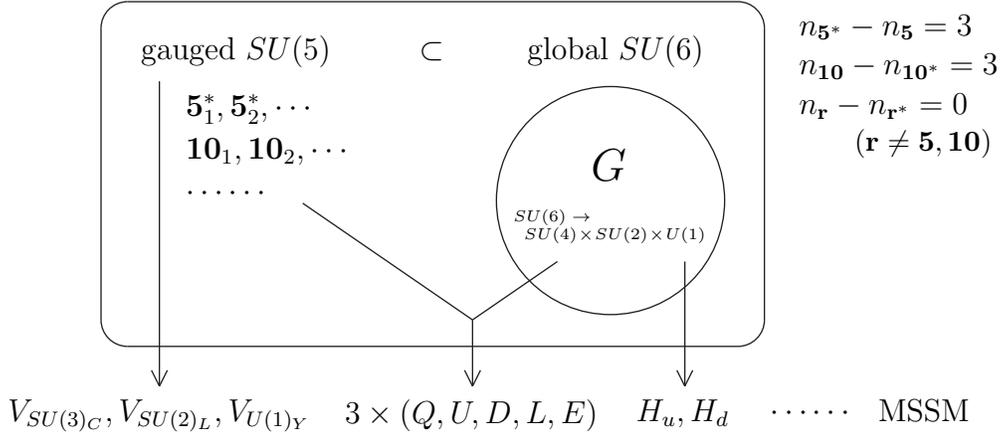
\begin{figure}[t]
\begin{center} 
\begin{picture}(310,170)(0,0)
  \CArc(10,50)(10,180,270)  \CArc(240,50)(10,270,360)
  \CArc(10,160)(10,90,180)  \CArc(240,160)(10,0,90)
  \Line(10,40)(240,40)  \Line(10,170)(240,170)
  \Line(0,50)(0,160)  \Line(250,50)(250,160)
  \Text(15,152)[l]{gauged $SU(5)$}  \Text(125,152)[]{$\subset$}
  \Text(32,130)[l]{${\bf 5}^*_1, {\bf 5}^*_2, \cdots$}
  \Text(32,114)[l]{${\bf 10}_1, {\bf 10}_2, \cdots$}
  \Text(32,98)[l]{$\cdots \cdots$}
  \CArc(192,95)(43,0,360)  \Text(228,152)[r]{global $SU(6)$}
  \Text(192,109)[]{\Large $G$}
  \Text(156,89)[l]{\tiny $SU(6) \rightarrow$}
  \Text(160,82)[l]{\tiny $SU(4) \!\times\! SU(2) \!\times\! U(1)$}
  \Text(264,161)[l]{$n_{{\bf 5}^*} - n_{\bf 5} = 3$}
  \Text(264,146)[l]{$n_{{\bf 10}} - n_{{\bf 10}^*} = 3$}
  \Text(264,131)[l]{$n_{{\bf r}} - n_{{\bf r}^*} = 0$}
  \Text(285,117)[l]{\small $({\bf r} \neq {\bf 5}, {\bf 10})$}
  \Line(22,140)(22,25) \Line(22,25)(19,31) \Line(22,25)(25,31)
  \Text(22,20)[t]{$V_{SU(3)_C}, V_{SU(2)_L}, V_{U(1)_Y}$}
  \Line(140,50)(172,72)  \Line(140,50)(76,94)
  \Line(140,50)(140,25) \Line(140,25)(137,31) \Line(140,25)(143,31)
  \Text(140,20)[t]{$3 \times (Q,U,D,L,E)$}
  \Line(220,72)(220,25) \Line(220,25)(217,31) \Line(220,25)(223,31)
  \Text(220,20)[t]{$H_u, H_d$}
  \Text(253,15)[l]{$\cdots \cdots$}
  \Text(294,16)[l]{MSSM}
\end{picture}
\caption{The basic picture of the theory in the 4D description.}
\label{fig:basic}
\end{center}
\end{figure}

We have depicted the basic picture of the theory in Fig.~\ref{fig:basic}. 
How can we realize this picture in explicit models?  It is not so 
straightforward to construct such models in the conventional 4D framework. 
In particular, it is not easy to find explicit gauge group and matter 
content for the $G$ sector having all the features described above.  (The 
difficulty increases if some of the relevant composite states are excited 
states of the $G$ sector.  We then cannot use beautiful exact results for 
${\cal N} = 1$ supersymmetric gauge theories~\cite{Intriligator:1995au}, 
which are applicable to lowest-lying modes.)  In our framework, however, 
this problem is in some sense ``bypassed.''  Suppose that the $G$ sector 
possesses a large 't~Hooft coupling, $\tilde{g}^2 \tilde{N} /16\pi^2 
\gg 1$.  In this case, the theory is so strongly coupled that the gauge 
theory description in terms of ``gluons'' and ``quarks'' does not make 
much sense.  Instead, in this parameter region, the theory is better 
specified by composite ``hadron'' states, which have a tower structure. 
For $\tilde{N} \gg 1$, these ``hadronic'' tower states are weakly 
coupled~\cite{'tHooft:1973jz}, and under certain circumstances 
they can be identified as the Kaluza-Klein (KK) states of a weakly 
coupled higher dimensional theory.  In particular, if the $G$ 
sector is quasi-conformal ($\tilde{g}$ evolves very slowly) 
above its dynamical scale, the corresponding higher dimensional 
theory is formulated in warped AdS spacetime truncated by 
branes~\cite{Maldacena:1997re,Arkani-Hamed:2000ds}.  In the next section 
we construct an explicit unified model in truncated 5D warped spacetime, 
which has all the features described in this section.  In practice, 
once we have a theory in higher dimensions, we can forget about the 
``original'' 4D picture for most purposes --- our higher dimensional 
theory is an effective field theory with which we can consistently 
calculate various physical quantities.  The theory does not require 
any more information than the gauge group, matter content, boundary 
conditions, and values of various parameters, to describe physics 
at energies below the cutoff scale $M_*$ ($\gg M_U$).

\section{Model}
\label{sec:model}

\subsection{Basic symmetry structure}
\label{subsec:symm}

Following the general picture presented in the previous section, we 
consider 5D warped spacetime truncated by two branes: the UV and IR 
branes.  The spacetime metric is given by
\begin{equation}
  ds^2 = e^{-2ky} \eta_{\mu\nu} dx^\mu dx^\nu + dy^2,
\label{eq:metric}
\end{equation}
where $y$ is the coordinate for the extra dimension and $k$ denotes the 
inverse curvature radius of the warped AdS spacetime.  The two branes 
are located at $y=0$ (the UV brane) and $y=\pi R$ (the IR brane).  This 
is the spacetime considered in Ref.~\cite{Randall:1999ee}, in which the 
AdS warp factor is used to generate the large hierarchy between the weak 
and the Planck scales by choosing the scales on the UV and IR branes 
to be the Planck and TeV scales, respectively ($kR \sim 10$).  Here 
we choose instead the UV-brane and IR-brane scales to be $k \approx 
(10^{17}-10^{18})~{\rm GeV}$ and $k' \equiv k\, e^{-\pi kR} \approx 
(10^{16}-10^{17})~{\rm GeV}$, respectively, so that the IR brane serves 
the role of breaking the unified symmetry.  (A more detailed discussion 
on the determination of the scales is provided in later subsections.) 
In this sense, we may loosely call the UV and IR branes the Planck 
and GUT branes, respectively. 

We consider supersymmetric unified gauge theory on this gravitational 
background.  We choose the gauge symmetry in the bulk to be $SU(6)$, 
corresponding to the global symmetry that the dynamical GUT-breaking 
sector possesses in the 4D description of the model.  The bulk 
$SU(6)$ gauge symmetry is broken to $SU(5) \times U(1)$ and $SU(4) 
\times SU(2) \times U(1)$ on the UV and IR branes, respectively, 
leaving an unbroken $SU(3) \times SU(2) \times U(1) \times U(1)$ gauge 
symmetry at low energies.  There are two ways to break a gauge symmetry 
on a brane: by boundary conditions and by the Higgs mechanism.  Let 
us first consider $SU(6) \rightarrow SU(5) \times U(1)$ on the UV 
brane.  If this breaking is caused by the Higgs mechanism, then in 
the corresponding 4D description the fundamental gauge symmetry 
of the theory is $SU(6)$, which is spontaneously broken to $SU(5) 
\times U(1)$ at a very high energy $E \gg M_U$.  In this case, we 
must introduce matter fields in representations of $SU(6)$, so that 
the standard $SU(5)$ embedding of matter fields~\cite{Georgi:1974sy} 
should be modified/extended.  On the other hand, if $SU(6) \rightarrow 
SU(5) \times U(1)$ on the UV brane is caused by boundary conditions, 
then in the corresponding 4D description only the $SU(5) \times 
U(1)$ subgroup of the global $SU(6)$ symmetry is explicitly gauged 
(see Fig.~\ref{fig:basic}), so that we can employ the standard $SU(5)$ 
embedding for matter fields.  We thus adopt the latter option to 
construct our minimal model here, although models based on the former 
option can also be accommodated in our framework.

\begin{figure}[t]
\begin{center}
  \input{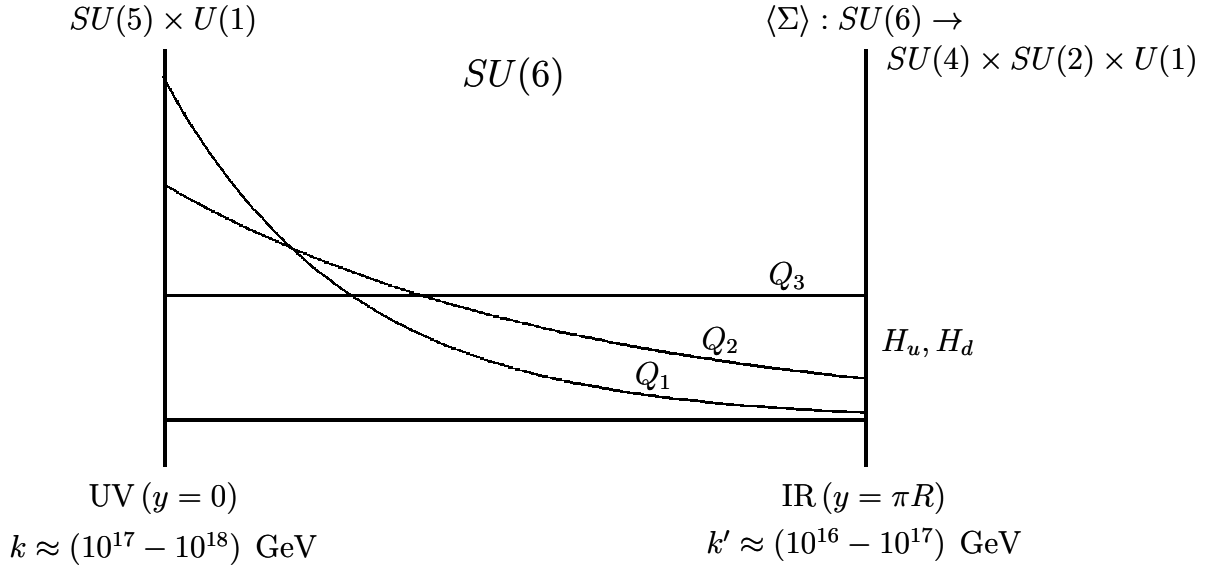}
\caption{A schematic picture of the model in 5D.}
\label{fig:5D}
\end{center}
\end{figure}
What about the symmetry breaking $SU(6) \rightarrow SU(4) \times SU(2) 
\times U(1)$ on the IR brane?  If we break $SU(6)$ to $SU(4) \times SU(2) 
\times U(1)$ by boundary conditions on the IR brane, the two massless 
Higgs doublets, whose existence is guaranteed by the general symmetry 
argument presented in the previous section, arise from extra-dimensional 
components of the bulk $SU(6)$ gauge fields.  This setup, however, leads 
to extra states lighter than $k' \approx (10^{16}-10^{17})~{\rm GeV}$ 
once matter fields are introduced in the bulk with the zero modes 
localized towards the UV brane (such matter fields are used to naturally 
explain the observed hierarchies in the fermion masses and mixings; 
see subsection~\ref{subsec:matter}).  These extra states generically 
do not fill complete $SU(5)$ representations and thus induce large 
threshold corrections for the standard model gauge couplings.  Large 
threshold corrections can be avoided if we judiciously choose boundary 
conditions for matter fields, but bulk $SU(6)$ gauge invariance then 
still requires complicated structure for the matter sector to reproduce 
the observed fermion masses and mixings.  These issues do not arise 
if the breaking $SU(6) \rightarrow SU(4) \times SU(2) \times U(1)$ 
is caused by the Higgs mechanism on the IR brane, as we will see later. 
We therefore adopt the Higgs breaking of $SU(6)$ on the IR brane. 
This completely determines the basic symmetry structure of our model, 
which is depicted in Fig.~\ref{fig:5D}.

\subsection{Gauge-Higgs sector and scales of the system}
\label{subsec:Higgs}

Let us start by describing the gauge-Higgs sector of the model.  Using 
4D $N=1$ superfield language, in which the gauge degrees of freedom 
are contained in $V(A_\mu, \lambda)$ and $\Phi(\phi+iA_5, \lambda')$, 
the boundary conditions for the 5D $SU(6)$ gauge supermultiplet are 
given by
\begin{eqnarray}
  && V:\: \left( \begin{array}{ccccc|c}
    (+,+) & (+,+) & (+,+) & (+,+) & (+,+) & (-,+) \\ 
    (+,+) & (+,+) & (+,+) & (+,+) & (+,+) & (-,+) \\ 
    (+,+) & (+,+) & (+,+) & (+,+) & (+,+) & (-,+) \\ 
    (+,+) & (+,+) & (+,+) & (+,+) & (+,+) & (-,+) \\ 
    (+,+) & (+,+) & (+,+) & (+,+) & (+,+) & (-,+) \\ \hline
    (-,+) & (-,+) & (-,+) & (-,+) & (-,+) & (+,+) 
  \end{array} \right),
\label{eq:bc-gauge-1} \\
  && \Phi:\: \left( \begin{array}{ccccc|c}
    (-,-) & (-,-) & (-,-) & (-,-) & (-,-) & (+,-) \\ 
    (-,-) & (-,-) & (-,-) & (-,-) & (-,-) & (+,-) \\ 
    (-,-) & (-,-) & (-,-) & (-,-) & (-,-) & (+,-) \\ 
    (-,-) & (-,-) & (-,-) & (-,-) & (-,-) & (+,-) \\ 
    (-,-) & (-,-) & (-,-) & (-,-) & (-,-) & (+,-) \\ \hline
    (+,-) & (+,-) & (+,-) & (+,-) & (+,-) & (-,-) 
  \end{array} \right),
\label{eq:bc-gauge-2}
\end{eqnarray}
where $+$ and $-$ represent Neumann and Dirichlet boundary conditions, 
respectively, and the first and second signs in parentheses represent 
boundary conditions at $y=0$ and $y=\pi R$, respectively.  These boundary 
conditions lead to $SU(6) \rightarrow SU(5) \times U(1)$ on the UV 
brane.  Since only $(+,+)$ components have zero modes, we obtain 4D 
$N=1$ $SU(5) \times U(1)$ gauge supermultiplets as massless fields at 
this point (coming from the upper left $5 \times 5$ block and the lower 
right element in $V$).  All the other KK modes have masses of order 
$\pi k'$ or larger.

The symmetry breaking $SU(6) \rightarrow SU(4) \times SU(2) \times 
U(1)$ on the IR brane is caused by the vacuum expectation value (VEV) 
of a field $\Sigma({\bf 35})$ localized to the IR brane, where the 
number in the parenthesis represents the transformation property under 
$SU(6)$.  We here consider that the $\Sigma$ field is strictly localized 
on the IR brane and has the following superpotential:
\begin{equation}
  {\cal L}_\Sigma = \delta(y-\pi R) 
    \biggl[ \int\! d^2\theta \Bigl( \frac{M}{2}\, {\rm Tr}(\Sigma^2) 
      + \frac{\lambda}{3}\, {\rm Tr}(\Sigma^3) \Bigr) 
    + {\rm h.c.} \biggr],
\label{eq:IR-Higgs}
\end{equation}
where the metric factor is absorbed into the normalization of 
the $\Sigma$ field.  (We will always absorb the metric factor into 
the normalizations of fields in similar expressions below, denoted 
by ${\cal L}$.)  The field $\Sigma$ is canonically normalized in 4D, 
so that natural values of the parameters $M$ and $\lambda$ are of 
order $M'_* = M_* e^{-\pi kR}$ and $4\pi$, respectively.  Here, 
$M_*$ is the cutoff scale of the 5D theory.  In general, the IR-brane 
potential for $\Sigma$ also has higher dimension terms suppressed 
by $M'_*$, in addition to Eq.~(\ref{eq:IR-Higgs}).  The presence 
of these terms, however, does not affect the qualitative conclusions 
of our paper.  Below, we assume that the parameter $M$ is a factor 
of a few smaller than its naive size, e.g. $M \sim k'$, to make 
our analysis better controlled.  In this case, the effect of higher 
dimension terms are expected to be suppressed even quantitatively.

The superpotential of Eq.~(\ref{eq:IR-Higgs}) has the following vacuum:
\begin{eqnarray}
  \langle \Sigma \rangle 
    &=& {\rm diag}\biggl(-\frac{2M}{\lambda}, -\frac{2M}{\lambda}, 
      \frac{M}{\lambda}, \frac{M}{\lambda}, \frac{M}{\lambda}, 
      \frac{M}{\lambda} \biggr),
\label{eq:Sigma-VEV}
\end{eqnarray}
where we have chosen $\lambda, M > 0$ without loss of generality. 
The VEV of Eq.~(\ref{eq:Sigma-VEV}) leads to $SU(6) \rightarrow 
SU(4) \times SU(2) \times U(1)$ on the IR brane, making a part of 
the $SU(5) \times U(1)$ gauge multiplet $V$ massive.  The remaining 
massless 4D $N=1$ gauge multiplet is that of $SU(3) \times SU(2) 
\times U(1) \times U(1)$, which we identify as the standard model 
gauge group with an extra $U(1)_X$: $SU(3)_C \times SU(2)_L \times 
U(1)_Y \times U(1)_X$.

An important aspect of the model is that the vacuum of 
Eq.~(\ref{eq:Sigma-VEV}) is a part of a continuum of vacua, 
which can easily be seen by studying the excitations.  Under 
the unbroken $SU(3)_C \times SU(2)_L \times U(1)_Y \times 
U(1)_X$ gauge symmetry, the $\Sigma$ field decomposes as
\begin{eqnarray}
  \Sigma &=& \Sigma_G({\bf 8}, {\bf 1})_{(0,0)}
    + \Sigma_W({\bf 1}, {\bf 3})_{(0,0)}
    + \Sigma_B({\bf 1}, {\bf 1})_{(0,0)}
\nonumber\\
  && {} + \Sigma_D({\bf 3}^*, {\bf 1})_{(1/3,2)}
    + \Sigma_{\bar{D}}({\bf 3}, {\bf 1})_{(-1/3,-2)}
    + \Sigma_L({\bf 1}, {\bf 2})_{(-1/2,2)}
    + \Sigma_{\bar{L}}({\bf 1}, {\bf 2})_{(1/2,-2)}
\nonumber\\
  && {} + \Sigma_X({\bf 3}, {\bf 2})_{(-5/6,0)}
    + \Sigma_{\bar{X}}({\bf 3}^*, {\bf 2})_{(5/6,0)}
    + \Sigma_{S}({\bf 1}, {\bf 1})_{(0,0)},
\label{eq:Sigma-comp}
\end{eqnarray}
where the numbers in parentheses represent the quantum numbers 
under $SU(3)_C \times SU(2)_L \times U(1)_Y \times U(1)_X$. 
The normalization of $U(1)_Y$ is chosen to match the conventional 
definition of hypercharge, while that of $U(1)_X$ is chosen, when 
matter fields are introduced, to match the conventional definition 
for the ``$U(1)_\chi$'' symmetry arising from $SO(10)/SU(5)$. 
Expanding the superpotential of Eq.~(\ref{eq:IR-Higgs}) around the 
vacuum, we find that all the components of $\Sigma$ obtain masses 
of order $M$ except for $\Sigma_X$, $\Sigma_{\bar{X}}$, $\Sigma_L$ 
and $\Sigma_{\bar{L}}$.  Among these four, the first two are absorbed 
into the massive $SU(5)/321$ gauge fields, but the last two remain 
as massless chiral superfields, which parameterize the continuous 
degeneracy of vacua.  This degeneracy is a consequence of the 
spontaneously broken $SU(6)$ symmetry, and the massless fields have 
the quantum numbers of a pair of Higgs doublets.  We thus identify 
these fields as the two Higgs doublets of the MSSM: $H_u$ and $H_d$. 

We have found that the gauge-Higgs sector of our model gives only 
the 4D $N=1$ gauge supermultiplet for $SU(3)_C \times SU(2)_L \times 
U(1)_Y \times U(1)_X$ and the two Higgs doublets $H_u$ and $H_d$ 
below the scale of order $M \sim k'$.  The $U(1)_X$ gauge symmetry 
can be broken at a scale somewhat below $k'$ by the Higgs mechanism. 
For example, we can introduce the superpotential on the UV brane 
\begin{equation}
  {\cal L}_X = \delta(y) 
    \biggl[ \int\! d^2\theta\, Y (X \bar{X} - \Lambda^2) 
    + {\rm h.c.} \biggr],
\label{eq:UV-X}
\end{equation}
where $Y$, $X$ and $\bar{X}$ are UV-brane localized chiral superfields 
that are singlet under $SU(5)$ and have charges of $0$, $10$ and 
$-10$ under $U(1)_X$, respectively.  The scale $\Lambda$ is that 
for $U(1)_X$ breaking, which may be generated by some other dynamics. 
The superpotential of Eq.~(\ref{eq:UV-X}) produces the VEVs $\langle 
X \rangle = \langle \bar{X} \rangle = \Lambda$, leading to $U(1)_X$ 
breaking at the scale $\Lambda$.  The nonvanishing VEV for $\bar{X}$ 
can also be used to generate small neutrino masses through the 
conventional seesaw mechanism, as we will see later.  This motivates
the values of the $X$ and $\bar{X}$ charges.

Various scales of our system -- the AdS inverse curvature radius 
$k$, the size of the extra dimension $R$, and the cutoff scale 
of the effective 5D theory $M_*$ -- are constrained by the scale 
of gauge coupling unification, the size of the unified gauge coupling 
$g_U \simeq 0.7$, and the value of the 4D (reduced) Planck scale 
$M_{\rm Pl} \simeq 2.4 \times 10^{18}~{\rm GeV}$.  In our warped 
5D theory, it is natural to consider that parameters in the bulk and 
on the IR brane obey naive dimensional analysis (at least roughly) 
while those on the UV brane do not, because the former represent 
strongly coupled $G$ dynamics while the latter represent the weakly 
coupled elementary sector.  Using naive dimensional analysis in 
higher dimensions~\cite{Chacko:1999hg}, we obtain the following 
Lagrangian for the graviton and the gauge fields:
\begin{equation}
  {\cal L} \approx \delta(y) 
    \Biggl[ \frac{\tilde{M}^2}{2} {\cal R}^{(4)}
      - \frac{1}{4 \tilde{g}^2} F^{\mu\nu} F_{\mu\nu} \Biggr]
    + \Biggl[ \frac{1}{2} \frac{M_*^3}{16\pi^3} {\cal R}^{(5)} 
      - \frac{1}{4} \frac{C M_*}{16\pi^3} F^{MN} F_{MN} \Biggr]
\label{eq:gravity-gauge}
\end{equation}
where ${\cal R}^{(4)}$ and ${\cal R}^{(5)}$ are the 4D and 5D Ricci 
curvatures, respectively, $M, N = 0,1,2,3,5$, and $C$ is a group 
theoretical factor, $C \simeq 6$.  This leads to the following relations:
\begin{eqnarray}
  \frac{1}{g_U^2} &\simeq& \frac{1}{\tilde{g}^2} 
    + \frac{C}{16\pi^2} \biggl(\frac{M_*}{\pi k} \biggr) \pi kR,
\label{eq:g_U}\\
  M_{\rm Pl}^2 &\simeq& \tilde{M}^2 
    + \frac{k^2}{16} \biggl(\frac{M_*}{\pi k} \biggr)^3 .
\label{eq:M_Pl}
\end{eqnarray}
Now, gauge coupling unification at $M_U \approx 10^{16}~{\rm GeV}$ 
implies that we should choose $M$ to be around this scale, and thus 
$k' = k\, e^{-\pi kR} \approx (10^{16}-10^{17})~{\rm GeV}$.  Then, 
choosing $M_*/\pi k$ to be a factor of a few, e.g. $M_*/\pi k \simeq 
(2\!\sim\!3)$, to make the higher dimensional description trustable, 
we obtain $k \simlt 10^{18}~{\rm GeV}$ from Eq.~(\ref{eq:M_Pl}) 
(and $\tilde{M}^2 > 0$).  We thus find that the scales of our 5D 
theory should be chosen as $k \approx (10^{17}-10^{18})~{\rm GeV}$ 
and $k' \approx (10^{16}-10^{17})~{\rm GeV}$, which implies 
$kR \sim 1$, with the cutoff scale $M_*$ a factor of a few larger 
than $\pi k$.  The UV-brane gauge coupling $\tilde{g}$ is then likely 
to be nonzero, implying that the elementary $SU(5)$ gauge field has 
nonvanishing tree-level kinetic terms in the 4D description.  In 
particular, this implies that elementary $SU(5)$ gauge interactions 
are likely to be weakly coupled at energies $E \gg M_U$.

\subsection{Matter sector and quark and lepton masses and mixings}
\label{subsec:matter}

Let us now include matter fields in the model.  In the 4D description 
of the theory, low-energy quark and lepton fields arise from mixtures 
of elementary states, which transform as ${\bf 10}$'s and ${\bf 5}^*$'s 
under the gauged $SU(5)$, and composite states of $G$, which form 
multiplets of the global $SU(6)$.  In the 5D theory, this situation is 
realized by introducing matter hypermultiplets in the bulk, which are 
representations of $SU(6)$, and by imposing $SU(6)$-violating boundary 
conditions on the UV brane.  We here present an explicit realization 
of this picture, leading to realistic phenomenology at low energies.

We begin by considering the structure of the matter sector for 
a single generation.  For quarks and leptons that are incorporated 
into the ${\bf 10}$ representation of $SU(5)$, $\{ Q, U, E \}$, 
we introduce a bulk hypermultiplet $\{ {\cal T}, {\cal T}^c \}$ 
transforming as ${\bf 20}$ under $SU(6)$:
\begin{eqnarray}
  {\cal T}({\bf 20})
    &=& {\bf 10}^{(+,+)}_{1}
      \oplus {\bf 10}^{*(-,+)}_{-1},
\label{eq:T} \\
  {\cal T}^c({\bf 20})
    &=& {\bf 10}^{*(-,-)}_{-1}
      \oplus {\bf 10}^{(+,-)}_{1},
\label{eq:Tc}
\end{eqnarray}
where ${\cal T}$ and ${\cal T}^c$ represent 4D $N=1$ chiral superfields 
that form a hypermultiplet in 5D.  (Our notation is such that 
``non-conjugated'' and ``conjugated'' chiral superfields have the 
opposite gauge quantum numbers; see e.g.~\cite{Arkani-Hamed:2001tb}. 
They have the same quantum numbers for ${\bf 20}$ of $SU(6)$ 
because ${\bf 20}$ is a (pseudo-)real representation.)  The 
right-hand-side of Eqs.~(\ref{eq:T},~\ref{eq:Tc}) shows the 
decomposition of ${\cal T}$ and ${\cal T}^c$ into representations 
of $SU(5) \times U(1)_X$ (in an obvious notation), as well as the 
boundary conditions imposed on each component (in the same notation 
as that in Eqs.~(\ref{eq:bc-gauge-1},~\ref{eq:bc-gauge-2})).  With 
these boundary conditions, the only massless state arising from 
$\{ {\cal T}, {\cal T}^c \}$ is ${\bf 10}_1$ of $SU(5) \times U(1)_X$ 
from ${\cal T}$, which we identify as the low-energy quarks and 
leptons $Q, U$ and $E$.

A bulk hypermultiplet $\{ {\cal H}, {\cal H}^c \}$ can generically 
have a mass term in the bulk, which is written as 
\begin{equation}
  S = \int\!d^4x \int_0^{\pi R}\!\!dy \, 
    \biggl[ e^{-3k|y|}\! \int\!d^2\theta\, 
    c_{\cal H}\, k\, {\cal H} {\cal H}^c + {\rm h.c.} \biggr],
\label{eq:bulk-mass}
\end{equation}
in the basis where the kinetic term is given by $S_{\rm kin} = \int\!d^4x 
\int\!dy\, [e^{-2k|y|} \int\!d^4\theta\, ({\cal H}^\dagger {\cal H} + 
{\cal H}^c {\cal H}^{c\dagger}) + \{ e^{-3k|y|} \int\!d^2\theta\, 
({\cal H}^c \partial_y {\cal H} - {\cal H} \partial_y {\cal H}^c)/2 + 
{\rm h.c.} \}]$~\cite{Marti:2001iw}.  The parameter $c_{\cal H}$ controls 
the wavefunction profile of the zero mode.  For $c_{\cal H} > 1/2$ 
($< 1/2$) the wavefunction of a zero mode arising from ${\cal H}$ is 
localized to the UV (IR) brane; for $c_{\cal H} = 1/2$ it is conformally 
flat.  (If a zero mode arises from ${\cal H}^c$, its wavefunction 
is localized to the IR (UV) brane for $c_{\cal H} > -1/2$ ($< -1/2$) 
and conformally flat for $c_{\cal H} = -1/2$.)  We choose these $c$ 
parameters to take values larger than about $1/2$ for matter fields. 
For these values of $c$ parameters, all the KK excited states of 
$\{ {\cal T}, {\cal T}^c \}$ have masses of order $\pi k'$ or larger, 
so that the $\{ {\cal T}, {\cal T}^c \}$ multiplet gives only the 
massless ${\bf 10}_1$ state below the energy scale of $k'$.

For quarks and leptons incorporated into the ${\bf 5}^*$ representation 
of $SU(5)$, $\{ D, L \}$, we introduce a bulk hypermultiplet $\{ {\cal F}, 
{\cal F}^c \}$ transforming as ${\bf 70}^*$ under $SU(6)$:
\begin{eqnarray}
  {\cal F}({\bf 70}^*)
    &=& {\bf 5}^{*(+,+)}_{-3}
      \oplus {\bf 10}^{*(-,+)}_{-1}
      \oplus {\bf 15}^{*(-,+)}_{-1}
      \oplus {\bf 40}^{*(-,+)}_{1},
\label{eq:F} \\
  {\cal F}^c({\bf 70})
    &=& {\bf 5}^{(-,-)}_{3}
      \oplus {\bf 10}^{(+,-)}_{1}
      \oplus {\bf 15}^{(+,-)}_{1}
      \oplus {\bf 40}^{(+,-)}_{-1},
\label{eq:Fc}
\end{eqnarray}
where the right-hand-side again shows the decomposition into 
representations of $SU(5) \times U(1)_X$, together with the boundary 
conditions imposed on each component.%
\footnote{Note that the signs $\pm$ for the boundary conditions in 
Eqs.~(\ref{eq:F},~\ref{eq:Fc}) represent the Neumann/Dirichlet boundary 
conditions in the interval $y: [0, \pi R]$.  In the orbifold picture, 
the boundary conditions of Eqs.~(\ref{eq:F},~\ref{eq:Fc}) can be 
obtained effectively as follows.  We prepare a hypermultiplet obeying 
the boundary conditions ${\cal F}({\bf 70}^*) = {\bf 5}^{*(+,+)}_{-3} 
\oplus {\bf 10}^{*(-,+)}_{-1} \oplus {\bf 15}^{*(-,+)}_{-1} \oplus 
{\bf 40}^{*(+,+)}_{1}$ and ${\cal F}^c({\bf 70}) = {\bf 5}^{(-,-)}_{3} 
\oplus {\bf 10}^{(+,-)}_{1} \oplus {\bf 15}^{(+,-)}_{1} \oplus 
{\bf 40}^{(-,-)}_{-1}$, where the first and second signs in the 
parentheses represent transformation properties under the reflection 
$y \leftrightarrow -y$ and $(y-\pi R) \leftrightarrow -(y - \pi R)$, 
respectively.  We then introduce a UV-brane localized chiral superfield 
transforming as ${\bf 40}_{-1}$ under $SU(5) \times U(1)_X$, and couple 
it to the ${\bf 40}^{*(+,+)}_{1}$ state from ${\cal F}({\bf 70}^*)$. 
This reproduces the boundary conditions of Eqs.~(\ref{eq:F},~\ref{eq:Fc}) 
in the limit that this coupling (brane mass term) becomes large.  (For 
the relation between a large brane mass term and the Dirichlet boundary 
condition, see e.g.~\cite{Nomura:2001mf}.)  The fact that the boundary 
conditions of Eqs.~(\ref{eq:F},~\ref{eq:Fc}) can be reproduced in 
the orbifold picture by taking a consistent limit guarantees their 
consistency.  In the 4D description, this corresponds to introducing 
only a ${\bf 5}^*_{-3}$ elementary state, which couples to 
a component of a $G$-invariant operator transforming as ${\bf 70}$ 
under the global $SU(6)$.  Similar remarks also apply to other 
fields, e.g. the $\{ {\cal N}, {\cal N}^c \}$ hypermultiplet 
in Eqs.~(\ref{eq:N},~\ref{eq:Nc}).}
With these boundary conditions, the only massless state arising from 
$\{ {\cal F}, {\cal F}^c \}$ is ${\bf 5}^*_{-3}$ of $SU(5) \times 
U(1)_X$ from ${\cal F}$, which we identify as the low-energy quarks 
and leptons $D$ and $L$.  All the KK excited states have masses of 
order $\pi k'$ or larger for $c_{\cal F} \simgt 1/2$.

The right-handed neutrino $N$ arises from a bulk hypermultiplet 
$\{ {\cal N}, {\cal N}^c \}$ transforming as ${\bf 56}$ of $SU(6)$:
\begin{eqnarray}
  {\cal N}({\bf 56})
    &=& {\bf 1}^{(+,+)}_{5}
      \oplus {\bf 5}^{(-,+)}_{3}
      \oplus {\bf 15}^{(-,+)}_{1}
      \oplus {\bf 35}^{(-,+)}_{-1},
\label{eq:N} \\
  {\cal N}^c({\bf 56}^*)
    &=& {\bf 1}^{(-,-)}_{-5}
      \oplus {\bf 5}^{*(+,-)}_{-3}
      \oplus {\bf 15}^{*(+,-)}_{-1}
      \oplus {\bf 35}^{*(+,-)}_{1}.
\label{eq:Nc}
\end{eqnarray}
The zero mode arises only from ${\bf 1}_5$ in ${\cal N}$, which is 
identified as the right-handed neutrino supermultiplet $N$. The other 
KK states are all heavier than of order $\pi k'$ for $c_{\cal N} 
\simgt 1/2$.

The Yukawa couplings for the quarks and leptons arise from IR-brane 
localized terms
\begin{equation}
  {\cal L}_{\rm Yukawa} = \delta(y-\pi R) 
    \biggl[ \int\! d^2\theta \Bigl( 
      y_{\cal T} {\cal T} {\cal T} \Sigma
      + y_{\cal F} {\cal T} {\cal F} \Sigma 
      + y_{\cal N} {\cal F} {\cal N} \Sigma \Bigr) 
    + {\rm h.c.} \biggr].
\label{eq:IR-Yukawa}
\end{equation}
(The Yukawa couplings also receive contributions from higher dimension 
terms as will be seen later in this subsection.)  Note that these 
interactions, as well as those in Eq.~(\ref{eq:IR-Higgs}), respect 
the usual $R$ parity of the MSSM, with $\Sigma$ even.

The interactions of Eq.~(\ref{eq:IR-Yukawa}) give the Yukawa 
couplings of the quark and lepton chiral superfields, $Q, U, D, L, E$ 
and $N$, with the Higgs doublets, $H_u$ and $H_d$, at low energies 
($W = QUH_u$, $QDH_d + LEH_d$ and $LNH_u$ from the first, second 
and third terms, respectively).  Recall that the two Higgs doublets 
of the MSSM, $H_u$ and $H_d$, arise from $\Sigma$ as pseudo-Goldstone 
chiral superfields of the broken $SU(6)$ symmetry.  For matter fields 
with $|c| > 1/2$, the Yukawa couplings receive suppressions due 
to the fact that the fields effectively feel only the IR brane 
(strong dynamics) or the UV brane (explicit breaking of $SU(6)$), 
both of which are needed to generate nonvanishing Yukawa couplings 
at low energies~\cite{Contino:2003ve}.  Then, considering that 
$y_{\cal T} \sim y_{\cal F} \sim y_{\cal N} = O(4\pi^2/M'_*)$ from 
naive dimensional analysis, we find that the low-energy Yukawa 
coupling $y$ arising from the IR-brane term $\int\!d^2\theta\, 
{\cal M}_1 {\cal M}_2 \Sigma$ (${\cal M}_1, {\cal M}_2 
= {\cal T}, {\cal F}, {\cal N}$) takes a value 
\begin{equation}
  y \approx 4\pi f_1 f_2\,
    \biggl( \frac{\pi k}{M_*} \biggr),
\label{eq:Yukawa-value}
\end{equation}
where $f_i \simeq (k'/k)^{|c_{{\cal M}_i}|-1/2}$ for $|c_{{\cal M}_i}| 
> 1/2$ and $f_i \simeq 1$ for $|c_{{\cal M}_i}| < 1/2$ ($i=1,2$); for 
$|c_{{\cal M}_i}| \simeq 1/2$, $f_i$ receives a logarithmic suppression, 
$f_i \simeq 1/(\ln(k/k'))^{1/2}$.  This allows us to explain the observed 
hierarchies of fermion masses and mixings by powers of $k'/k = e^{-\pi kR} 
= O(0.1)$, by choosing different values of $c_{\cal T}, c_{\cal F}$ and 
$c_{\cal N}$ for different generations.  This is similar to the situation 
where the hierarchies are explained by overlaps of matter and Higgs 
wavefunctions~\cite{Gherghetta:2000qt,Hebecker:2002re}, although 
in the present setup the low-energy Yukawa couplings are also suppressed 
for $c_{{\cal M}_i} < -1/2$, where apparent overlaps between matter 
and Higgs fields are large, due to the pseudo-Goldstone boson nature 
of the Higgs doublets.  This opens the possibility of localizing the 
first two generations to the IR brane, rather than to the UV brane 
as we will do shortly, to generate the observed hierarchies of fermion 
masses and mixings.

The right-handed neutrino superfield $N$ can obtain a large mass 
term through the UV-brane operator
\begin{equation}
  {\cal L}_N = \delta(y) 
    \biggl[ \int\! d^2\theta\, \frac{\eta}{2} \bar{X} N^2 
    + {\rm h.c.} \biggr],
\label{eq:UV-N}
\end{equation}
where $\bar{X}$ is a $U(1)_X$-breaking field, having the VEV 
$\langle \bar{X} \rangle = \Lambda$ (see Eq.~(\ref{eq:UV-X})). 
This gives a small mass for the observed left-handed neutrino 
through the conventional seesaw mechanism~\cite{Seesaw}.%
\footnote{An alternative possibility to generate a small neutrino 
mass is to strongly localize the $N$ field to the IR brane by taking 
$c_{\cal N} \ll -1/2$, in which case the neutrino Yukawa coupling 
is strongly suppressed and we can obtain a small Dirac neutrino 
mass.  The scale of the neutrino mass, however, is unexplained 
in this case.}

It is rather straightforward to generalize the analysis so far 
to the case of three generations.  We simply introduce a set of 
bulk hypermultiplets $\{ {\cal T}, {\cal T}^c \}$, $\{ {\cal F}, 
{\cal F}^c \}$ and $\{ {\cal N}, {\cal N}^c \}$ for each generation. 
The couplings $y_{\cal T}$, $y_{\cal F}$ and $y_{\cal N}$ in 
Eq.~(\ref{eq:IR-Yukawa}) and $\eta$ in Eq.~(\ref{eq:UV-N}) then 
become $3 \times 3$ matrices.  We assume that there is no special 
structure in these matrices, so that all the elements in $y_{\cal T}$, 
$y_{\cal F}$ and $y_{\cal N}$ are of order $4\pi^2/M'_*$, suggested 
by naive dimensional analysis.  The observed fermion masses and 
mixings, however, can still be reproduced through the dependence of 
the low-energy Yukawa couplings on the values of bulk hypermultiplet 
masses $c_{\cal T}, c_{\cal F}$ and $c_{\cal N}$.  Let us take, for 
example, the bulk masses to be
\begin{equation}
  c_{{\cal T}_1} \simeq \frac{5}{2}, \quad 
  c_{{\cal T}_2} \simeq \frac{3}{2}, \quad 
  c_{{\cal T}_3} \simeq \frac{1}{2}, \quad
  c_{{\cal F}_1} \simeq c_{{\cal F}_2} \simeq 
  c_{{\cal F}_3} \simeq \frac{3}{2}, \quad
  c_{{\cal N}_1} \simeq c_{{\cal N}_2} \simeq 
  c_{{\cal N}_3} \simeq \frac{1}{2}.
\label{eq:c-flavor}
\end{equation}
Then, taking $M_*/\pi k$ to be a factor of a few, e.g. $2\!\sim\!3$, 
we obtain the following low-energy Yukawa matrices from 
Eq.~(\ref{eq:Yukawa-value}):
\begin{equation}
  y_u \approx 
  \pmatrix{
     \epsilon^4 & \epsilon^3 & \epsilon^2 \cr
     \epsilon^3 & \epsilon^2 & \epsilon   \cr
     \epsilon^2 & \epsilon   & 1          \cr
  },
\quad
  y_d \approx y_e^T \approx
  \epsilon
  \pmatrix{
     \epsilon^2 & \epsilon^2 & \epsilon^2 \cr
     \epsilon   & \epsilon   & \epsilon   \cr
     1          & 1          & 1          \cr
  },
\quad
  y_\nu \approx
  \epsilon
  \pmatrix{
     1 & 1 & 1 \cr
     1 & 1 & 1 \cr
     1 & 1 & 1 \cr
  },
\label{eq:y-values}
\end{equation}
where $y_u$, $y_d$, $y_e$ and $y_\nu$ are defined in the low-energy 
superpotential by
\begin{equation}
  W = (y_u)_{ij} Q_i U_j H_u + (y_d)_{ij} Q_i D_j H_d
    + (y_e)_{ij} L_i E_j H_d + (y_\nu)_{ij} L_i N_j H_u,
\label{eq:y-def}
\end{equation}
with $i,j, = 1,2,3$, and
\begin{equation}
  \epsilon \equiv \frac{k'}{k} \simeq \frac{1}{20}
  \quad {\rm for} \,\,\, kR \simeq 1.
\label{eq:epsilon}
\end{equation}
Together with a structureless Majorana mass matrix for the 
right-handed neutrinos, $M_N = \eta \langle \bar{X} \rangle$, 
the Yukawa matrices of Eq.~(\ref{eq:y-values}) well reproduces 
gross features of the observed quark and lepton masses and 
mixings~\cite{Hall:1999sn}.  It is straightforward to make 
further refinements on this basic picture; for example, we can 
make $c_{{\cal F}_1}$ somewhat larger than $3/2$ to better reproduce 
down-type quark and charged lepton masses, as well as the neutrino 
mixing angles $\theta_{12}$ and $\theta_{13}$.  A schematic picture 
for the zero-mode wavefunctions (for the $\{ {\cal T}, {\cal T}^c \}$ 
multiplets) is depicted in Fig.~\ref{fig:5D}.

Unwanted $SU(5)$ mass relations for the first two generation 
fermions can be avoided by using higher dimension operators, e.g. 
of the form ${\cal L} \sim \delta(y-\pi R) \int\!d^2\theta\, {\cal T} 
{\cal F} \Sigma^2$.  (Violation of $SU(5)$ relations may also come 
from $SU(5)$-violating mixings between the matter zero modes and 
the corresponding KK excited states, arising from the IR-brane 
terms of Eq.~(\ref{eq:IR-Yukawa}) through the $\Sigma$ VEV.)  Since 
the effects are higher order in $\langle \Sigma \rangle/M'_*$, 
which we assume somewhat small, $O(1)$ violation in the Yukawa 
coupling requires a somewhat suppressed coefficient for the leading 
$SU(5)$-invariant piece coming from Eq.~(\ref{eq:IR-Yukawa}). 
A realistic pattern for the fermion masses and mixings can be 
obtained if (only) the 22 element of the $y_{\cal F}$ matrix 
is somewhat suppressed~\cite{Georgi:1979df}.

The three generation model allows IR-brane operators of the form 
${\cal L} \sim \delta(y-\pi R) \int\!d^2\theta\, \epsilon^{ij} 
{\cal T}_i {\cal T}_j$, where the antisymmetry in the generation 
indices $i,j$ arises from the pseudo-real nature of the ${\bf 20}$ 
representation.  The existence of these operators, however, does 
not significantly affect predictions of the model.

To summarize, we have obtained an $SU(3)_C \times SU(2)_L \times 
U(1)_Y \times U(1)_X$ gauge theory below the scale of $M \sim k'$, 
with three generations of matter, $Q, U, D, L, E$ and $N$, and 
two Higgs doublets, $H_u$ and $H_d$.  The Yukawa couplings of 
Eq.~(\ref{eq:y-def}) are obtained with realistic patterns for quark 
and lepton masses and mixings.  The $U(1)_X$ gauge symmetry is 
spontaneously broken at the scale $\Lambda$, somewhat below $k'$, 
giving masses to the right-handed neutrino superfields of order 
$\Lambda$.  We thus have the complete MSSM, supplemented by seesaw 
neutrino masses, below the unification scale $\sim k'$.  We emphasize 
that the successes of our model depend only on its basic features, 
such as the symmetry structure and locations of fields.  They are 
thus quite robust.  For example, the existence of higher dimension 
operators in the IR-brane potential, e.g. terms of the form ${\rm 
Tr}(\Sigma^n)$ ($n$: integers $>3$) added to Eq.~(\ref{eq:IR-Higgs}), 
does not destroy these successes.

\subsection{Gauge coupling unification and proton decay}
\label{subsec:analysis}

In this subsection, we present a study on proton decay and gauge 
coupling unification in our model, to demonstrate that it can 
accommodate realistic phenomenology at low energies.  In this 
subsection we consider matter configurations such that lighter 
generations are localized more towards the UV brane, as in the 
example of Eq.~(\ref{eq:c-flavor}).

We first note that the terms in Eqs.~(\ref{eq:IR-Yukawa}) introduce, 
through the VEV of $\Sigma$, $SU(5)$-violating mass splittings 
into the KK towers for the matter fields: $\{ {\cal T}, {\cal T}^c \}$, 
$\{ {\cal F}, {\cal F}^c \}$ and $\{ {\cal N}, {\cal N}^c \}$. 
These splittings, in turn, give threshold corrections to gauge 
coupling unification.  Similar corrections also arise from the 
gauge KK towers.  We expect, however, that these corrections are 
not large.  Using the AdS/CFT correspondence, we estimate the 
size of the corrections to be of order $(C/16\pi^2)(M_*/\pi k)$ 
for $1/g_a^2$, where $g_a$ are the 4D gauge couplings.  Moreover, 
if the value of $\langle \Sigma \rangle$ (and thus $M$) is somewhat 
suppressed compared with its naive size of $M'_*/4\pi$, as we assume 
here, the threshold corrections receive additional suppressions 
of $O(4\pi \langle \Sigma \rangle/M'_*)$ because the spectrum 
of the KK towers becomes $SU(5)$ symmetric for $4\pi \langle \Sigma 
\rangle/M'_* \ll 1$.  The contributions from tree-level IR-brane 
operators, such as $\int\!d^2\theta\, \Sigma {\cal W}^\alpha 
{\cal W}_\alpha$, are also sufficiently small, of order $C/16\pi^2$ 
for $1/g_a^2$ with an additional suppression of $O(4\pi \langle 
\Sigma \rangle/M'_*)$ for small $\langle \Sigma \rangle$.

Another important issue in supersymmetric unified theories is 
dimension five proton decay caused by low-energy operators of the 
form $W \sim QQQL$, $UUDE$.  There are two independent sources for 
these operators: tree-level operators existing at the gravitational 
scale and operators generated by the GUT (breaking) dynamics.  In 
our theory, the former correspond to tree-level operators ${\bf 10}_1 
{\bf 10}_1 {\bf 10}_1 {\bf 5}^*_{-3} \supset QQQL$, $UUDE$ located 
on the UV brane, where the subscripts on ${\bf 10}_1 \subset {\cal T}$ 
and ${\bf 5}^*_{-3} \subset {\cal F}$ denote the $U(1)_X$ charges. 
While the coefficients of these operators are suppressed by the 
fundamental scale $M_*$, which is larger than the unification scale, 
it is still problematic, especially because we do not have any 
Yukawa suppressions in the coefficients.  Therefore, to suppress 
these contributions, we impose a discrete $Z_{4,R}$ symmetry on the 
theory, whose charge assignment is given in Table~\ref{table:Z4R} 
(in the normalization that the $R$ charge of the superpotential 
is $2$).
\begin{table}
\begin{center}
\begin{tabular}{|c|cc|c|cccccc|ccc|}  \hline 
  & $V$ & $\Phi$ & $\Sigma$ & 
    ${\cal T}$ & ${\cal T}^c$ & ${\cal F}$ & 
    ${\cal F}^c$ & ${\cal N}$ & ${\cal N}^c$ & 
    $Y$ & $X$ & $\bar{X}$ \\ \hline
  $Z_{4,R}$ & 0 & 0 & 0 & 1 & 1 & 1 & 1 & 1 & 1 & 
    2 & 0 & 0 \\ \hline
\end{tabular}
\end{center}
\caption{$Z_{4,R}$ charges for fields.}
\label{table:Z4R}
\end{table}
As is clear from the terms in Eq.~(\ref{eq:IR-Higgs}), this symmetry 
should be broken on the IR brane, i.e. broken by the dynamics of 
the GUT breaking sector $G$.  This can be incorporated by introducing 
a spurion chiral superfield $\phi$ with $\langle \phi \rangle \sim 
M'_*/4\pi$ on the IR brane, whose $Z_{4,R}$ charge is $+2$, or 
equivalently introducing fields $\phi$ and $\bar{\phi}$ of $Z_{4,R}$ 
charges $+2$ and $-2$ with the superpotential giving the VEVs for 
these fields.  This introduces ``$O(1)$'' breaking of $Z_{4,R}$ on 
the IR brane, keeping $Z_{4,R}$ invariance for the UV-brane terms.%
\footnote{The $Z_{4,R}$ symmetry can be gauged in 5D if we cancel 
the discrete $Z_{4,R}$-$SU(5)^2$ anomaly by the Green-Schwarz 
mechanism~\cite{Green:1984sg}, by introducing a singlet field $S$ 
that transforms nonlinearly under $Z_{4,R}$ and couples to the 
$SU(5)$ gauge kinetic term on the UV brane.  We consider that the 
$S$ field appears only in front of the kinetic term of the $SU(5)$ 
gauge superfields, and not in UV-brane superpotential terms.  Such 
terms would potentially induce dimension five proton decay, although 
they are suppressed in a certain (broad) region for the $S$ VEV.}

After killing the UV-brane operators, the low-energy dimension five 
proton decay operators can still be generated through strong $G$ 
dynamics, since the $Z_{4,R}$ symmetry is spontaneously broken by 
this dynamics.  One source is tree-level dimension five operators 
on the IR brane.  These operators, however, receive suppressions 
of order the Yukawa couplings in 4D, because the wavefunctions for 
light generation matter are suppressed on the IR brane due to the 
bulk hypermultiplet masses, and so are not particularly dangerous. 
(In the 4D picture, these suppressions arise from small mixings 
between the elementary and composite matter states for light 
generations.)  The only potentially dangerous contribution to 
dimension five proton decay in our model then comes from the exchange 
of the colored triplet Higgsinos -- composite states of $G$ arising 
as components of $\Sigma$ -- because the mass of these states can 
be smaller than $M'_*$.  To suppress this contribution, we can simply 
raise the mass of the colored triplet Higgsino states compared with 
the unification scale; in fact, the mass is expected to be larger 
than the GUT breaking VEV because the coupling $\lambda$ in 
Eq.~(\ref{eq:IR-Higgs}) is naturally of order $4\pi$.  Note that 
because of the existence of threshold corrections from KK towers 
to gauge coupling unification, there is no tight relation between 
the mass of the triplet Higgsinos and the low-energy values of 
the gauge couplings, which excluded the minimal SUSY $SU(5)$ GUT 
in 4D~\cite{Murayama:2001ur}.

\subsection{Supersymmetry breaking}
\label{subsec:SUSY-breaking}

Our model can be combined with almost any supersymmetry breaking 
scenario.  If the mediation scale of supersymmetry breaking is 
lower than the unification scale, there are essentially no particular 
implications from our theory on the pattern of supersymmetry breaking. 
On the other hand, if the mediation scale is higher, there can be 
interesting implications, e.g., on the flavor structure of supersymmetry 
breaking masses.  For example, if the supersymmetry breaking sector 
is localized on the IR brane, i.e. arises as a result of the dynamics 
of $G$, the third generation superparticles (presumably only the 
ones coming from the ${\bf 10}$ representation of $SU(5)$) can 
have different masses than the lighter generation superparticles, 
which receive universal masses from the gauginos through loop 
corrections~\cite{Kitano:2006}.%
\footnote{We thank R.~Kitano for discussions on this issue.}
These are consequences of our way of generating hierarchies 
in fermion masses and mixings.

The supersymmetric mass (the $\mu$ term) and supersymmetry-breaking 
masses (the $\mu B$ term and non-holomorphic scalar squared masses) 
for the Higgs doublets are both generated through supersymmetry 
breaking.  In the case that the supersymmetry breaking sector is 
localized on or directly communicates with the IR brane, these masses 
are generated through IR-brane operators of the form, ${\cal L} 
\sim \delta(y-\pi R) \int\!d^2\theta\, \{ Z M \Sigma^2/M'_* + 
Z \Sigma^3/M'_*\} + {\rm h.c.}$ and $\delta(y-\pi R) \int\!d^4\theta\, 
\{ (Z+Z^\dagger)\Sigma^\dagger \Sigma/M'_* + Z^\dagger Z (\Sigma^2 
+ \Sigma^{\dagger 2})/M_*^{\prime 2} + Z^\dagger Z \Sigma^\dagger 
\Sigma/M_*^{\prime 2} \}$, where $Z$ is a chiral superfield responsible 
for supersymmetry breaking, $\langle Z \rangle = \theta^2 F_Z$, 
and we have omitted $O(1)$ coefficients.  These operators produce 
supersymmetry breaking terms in the $\Sigma$ potential, which lead 
to a slight shift of the vacuum from Eq.~(\ref{eq:Sigma-VEV}) and 
consequently generate weak scale masses for components of $H_u$ 
and $H_d$.  The generated masses respect the relation
\begin{equation}
  \mu B = \left||\mu|^2 + m_{H_u}^2 \right|,
\qquad
  m_{H_u}^2 = m_{H_d}^2,
\label{eq:Higgs-1}
\end{equation}
reflecting the fact that the scalar potential for $\Sigma$ still 
has a global $SU(6)$ symmetry, where $m_{H_u}^2$ and $m_{H_d}^2$ are 
non-holomorphic supersymmetry breaking squared masses for $H_u$ and 
$H_d$, and we have taken the phase convention that $\mu B > 0$.  Note 
that, unlike the case where the Higgs fields are non pseudo-Goldstone 
fields~\cite{Giudice:1988yz}, the K\"ahler potential terms of the 
form $\delta(y-\pi R) \int\!d^4\theta\, \Sigma^2$, $\delta(y-\pi R) 
\int\!d^4\theta\, \{ Z^\dagger \Sigma^2/M'_* + {\rm h.c.} \}$ and 
$\delta(y-\pi R) \int\!d^4\theta\, \{ Z^\dagger Z \Sigma^2/M_*^{\prime 2} 
+ {\rm h.c.} \}$ do not produce a weak scale $\mu$ term; we need 
supersymmetry breaking interactions for $\Sigma$, generated by 
superpotential terms or $\delta(y-\pi R) \int\!d^4\theta\, 
(Z+Z^\dagger)\Sigma^\dagger \Sigma/M'_*$. 

An interesting case arises if supersymmetry is broken in a hidden 
sector that does not have direct interactions with the GUT breaking 
sector.  In this case, the Higgs sector supersymmetry breaking 
parameters arise through gravitational effects and obey the 
tighter relation
\begin{equation}
  \mu B = |\mu|^2, \qquad m_{H_u}^2 = m_{H_d}^2 = 0.
\label{eq:Higgs-2}
\end{equation}
In the language of the compensator formalism (see 
e.g.~\cite{Randall:1998uk}), these terms arise from ${\cal L} \sim 
\delta(y-\pi R) \int\!d^2\theta\, \phi (M/2) \Sigma^2 + {\rm h.c.}$, 
where $\phi = 1 + \theta^2 m_{3/2}$ is the compensator field 
with $m_{3/2}$ the gravitino mass.  (Here, we have assumed that 
supersymmetry breaking in the compensator field is not canceled 
by the conformal dynamics of the GUT breaking sector.)  The relations 
of Eqs.~(\ref{eq:Higgs-1},~\ref{eq:Higgs-2}) hold at the unification 
scale of $O(k')$, so that their connections to low energy parameters 
must involve renormalization group effects between the unification 
and the weak scales.  It is also possible that there are 
additional contributions to supersymmetry breaking parameters, 
e.g. $m_{H_u}^2$ and $m_{H_d}^2$, in addition to the ones in 
Eqs.~(\ref{eq:Higgs-1},~\ref{eq:Higgs-2})

Alternatively, the $\mu$ and $\mu B$ terms may be generated below 
the unification scale.  For example, they may be generated associated 
with the dynamics of $U(1)_X$ breaking~\cite{Hall:2002up}.  In this 
case there is no trace in the Higgs sector parameters that the Higgs 
fields are pseudo-Goldstone fields of the GUT breaking dynamics.

\section{Other Theories: GUT Engineering on the IR Brane}
\label{sec:other}

So far, we have considered a theory in which the lightness of the 
two Higgs doublets is understood by the pseudo-Goldstone mechanism 
associated with the dynamics of GUT breaking.  As we have seen, this 
can be elegantly implemented in our framework by considering the 
bulk $SU(6)$ gauge symmetry, broken to $SU(5) \times U(1)$ and 
$SU(4) \times SU(2) \times U(1)$ on the UV and IR branes, respectively. 
The mass of the light Higgs doublets is protected from the existence 
of explicit breaking by localizing these fields ($\subset \Sigma$) 
on the IR brane, which is geometrically separated from the UV brane 
where explicit breaking of $SU(6)$ resides.  In the 4D description 
of the model, the global $SU(6)$ symmetry of the GUT breaking sector 
is understood as a ``flavor'' symmetry of this sector, and the extreme 
suppression of explicit symmetry breaking effects in the $\Sigma$ 
potential comes from the fact that $\Sigma$ is a composite field, 
with the corresponding operator having a (very) large canonical mass 
dimension.  An interesting thing about our construction is that it 
allows us to implement these mechanisms in simple and controllable 
ways in effective field theory, giving a simple and calculable unified 
theory above $M_U$ in which the lightness of the Higgs doublets is 
understood by a symmetry principle.

Let us now consider if we can construct simpler theories in our warped 
space framework.  Suppose we consider a supersymmetric $SU(5)$ gauge 
theory in the 5D warped spacetime of Eq.~(\ref{eq:metric}), and suppose 
that the bulk $SU(5)$ gauge symmetry is broken to the $SU(3)_C \times 
SU(2)_L \times U(1)_Y$ subgroup on the IR brane by boundary conditions. 
In this case, the doublet Higgs fields may be light without being 
accompanied by their triplet partners if they propagate in the 
bulk with appropriate boundary conditions imposed at the GUT breaking 
brane~\cite{Kawamura:2000ev,Hall:2001pg}, or if they are simply 
located on that brane~\cite{Hebecker:2001wq}.  In the 4D description, 
however, this seems to be simply ``postulating'' particular dynamics 
of the GUT breaking sector that splits the mass of the doublet components 
from that of the triplet partners, and it is not clear if this can be 
regarded as a ``solution'' to the doublet-triplet splitting problem. 
For example, we have a continuous parameter, the tree-level mass of the 
Higgs doublets on the IR brane, that has to be chosen to be very small 
to achieve the splitting.  The situation may be better if this parameter 
is forbidden by a symmetry, e.g. an $R$ symmetry~\cite{Hall:2001pg}. 
This symmetry may be imposed as a global symmetry in 5D, but in that 
case it is not entirely clear if such a symmetry is preserved by strong 
$G$ dynamics (5D quantum gravity effects).  To avoid this and to give 
a non-trivial meaning to the symmetry in the context of gauge/gravity 
duality, we can gauge the symmetry in higher dimensions (although it 
can still be broken on the UV brane, eliminating the existence of the 
corresponding gauge field in 4D).  In this case, anomaly cancellation 
conditions become an issue, and we find that for a continuous $U(1)_R$ 
or a discrete $Z_{4,R}$ symmetry (with the charge assignment given 
by $V_{SU(5)}(0)$, $T_{\bf 10}(1)$, $F_{{\bf 5}^*}(1)$, $N_{\bf 1}(1)$, 
$H_{\bf 5}(0)$, $\bar{H}_{{\bf 5}^*}(0)$, assuming the MSSM matter 
content at low energies) we need to cancel the low energy anomalies 
via the Green-Schwarz mechanism~\cite{Green:1984sg}.  This requires the 
introduction of a singlet field $S$ on the IR brane which transforms 
nonlinearly under the $R$ symmetry and couples to the $SU(3)_C$, $SU(2)_L$ 
and $U(1)_Y$ gauge kinetic terms with appropriate coefficients.  (Anomaly 
transmission across the bulk~\cite{Callan:1984sa} may also be necessary 
to make the full 5D theory anomaly free, depending on the symmetry 
and matter content.)  We assume that the $S$ field appears only in 
front of the gauge kinetic terms, and not in IR-brane superpotential 
terms, so that a large mass term for the Higgs doublets is not 
regenerated.  (This may naturally occur in a UV theory in the absence 
of other gauge groups.)  Note that, in the 4D description, this 
setup corresponds to the situation where the $R$-$SU(5)^2$ anomaly 
is canceled between the elementary-field and $G$-sector contributions.%
\footnote{We can show that this construction is not available in 
a 4D SUSY GUT theory where the GUT-breaking (Higgs) sector does not 
give tree-level contributions to the low energy anomalies.  Assuming 
the MSSM matter content below the unification scale, with the $U(1)_R$ 
charges given by $V_{321}(0)$, $Q(1)$, $U(1)$, $D(1)$, $L(1)$, $E(1)$, 
$H_u(0)$ and $H_d(0)$, we find the low-energy $U(1)_R$-$SU(3)_C^2$, 
$U(1)_R$-$SU(2)_L^2$ and $U(1)_R$-$U(1)_Y^2$ anomalies to be $3$, 
$1$ and $-3/5$, respectively, which cannot be matched to high 
energy theories, where these anomalies arise as a $U(1)_R$-$SU(5)^2$ 
anomaly and are thus universal.  (Here, the $SU(5)$ normalization 
is employed for the $U(1)_Y$ charges.)  This implies that $U(1)_R$ 
should either be spontaneously broken, or there is explicit 
$SU(5)$-violating physics in the effective field theory.  In our case, 
this conclusion can be avoided because the (dynamical) GUT-breaking 
sector carries the $U(1)_R$-$SU(5)^2$ anomaly, a part of which can 
be manifested as Green-Schwarz terms at low energies.} 
In this setup, doublet-triplet splitting seems ``natural,'' at least 
in the higher dimensional picture.  Thus, while the theory with the 
$R$ symmetry still seems to correspond to a particular choice of GUT 
breaking dynamics in the 4D description, we may say that the theory 
does not have the problem of doublet-triplet splitting.%
\footnote{An interesting feature of this class of theories is 
that the low energy theory contains an axion field $S$ that 
couples to the QCD gauge fields with the decay constant of order 
the unification scale.  This can be used to solve the strong $CP$ 
problem~\cite{Peccei:1977hh}, although the initial amplitude of 
this field in the early universe must be (accidentally) small to 
avoid the cosmological difficulty of overclosing the universe.}
After all, the ``formulation'' of the doublet-triplet splitting problem 
may have to be changed in the large 't~Hooft coupling regime, where 
physics is specified by the ``hadronic'' quantities, i.e. matter 
content, location, and boundary conditions in higher dimensions.%
\footnote{If we break the bulk $SU(5)$ gauge symmetry by boundary 
conditions at the UV brane, it leads to a theory which is interpreted 
as an $SU(3)_C \times SU(2)_L \times U(1)_Y$ gauge theory in the 4D 
description.  The doublet-triplet splitting problem does not arise 
as the theory is not unified, and yet the successful unification of 
gauge couplings arises at the leading-log level in the limit that 
the tree-level gauge kinetic terms on the UV brane are small.  This 
corresponds in the 4D description that the 321 gauge couplings 
at the unification scale are dominated by the asymptotically non-free 
contribution from a strong sector that has a global $SU(5)$ symmetry, 
of which the $SU(3) \times SU(2) \times U(1)$ subgroup is gauged 
and identified as the low-energy 321 gauge group.  While this theory 
is somewhat outside the framework described in this paper, it is 
interesting on its own.}

In these respects, our framework offers many possible ways to 
address the problems of conventional 4D SUSY GUTs.  For example, we 
can again consider a 5D $SU(5)$ gauge theory in the warped spacetime 
of Eq.~(\ref{eq:metric}), but then break the bulk $SU(5)$ by the 
VEV of a chiral superfield located on the IR brane, generated by 
an appropriate IR-brane superpotential.  Then, if this superpotential 
does not have the problem of doublet-triplet splitting, e.g. by having 
the form of missing partner type models~\cite{Masiero:1982fe}, then 
we may say that the problem has been solved.  (As discussed before, 
it is better if the IR-brane superpotential is protected by a (discrete) 
symmetry; otherwise, it would correspond to ``artificially'' choosing 
the dynamics of the GUT breaking sector.  In practice, this may be 
difficult, since we cannot use the non-universal Green-Schwarz terms 
on the IR brane because the GUT breaking there is due to the Higgs 
mechanism.  We do not pursue this issue further here.)  An advantage 
of this approach over conventional 4D model building is that we need 
not care about physics above the unification scale when engineering 
GUT breaking physics, i.e. the GUT-breaking Higgs content and 
superpotential.  In the conventional 4D SUSY GUT framework, theories 
solving the doublet-triplet and/or dimension five proton decay 
problems often have too large matter content, leading to the problem 
of the unified gauge coupling hitting a Landau pole (well) below 
the gravitational/Planck scale.  In our case, all (possibly large) 
multiplets located on the IR brane correspond to composite fields 
of the GUT breaking dynamics in the 4D description, and do not 
contribute to the running of the unified gauge coupling above the 
unification scale, $M_U \sim k'$ (see e.g.~\cite{Goldberger:2002cz}). 
Potential complication of this sector may also not bother us, because 
it is the result of ``dynamics'' of the GUT breaking sector.  We note 
that this makes the extension to $SO(10)$ unified theories trivial 
--- we can break $SO(10)$ on the IR brane by arbitrary combinations 
of boundary condition and Higgs breakings with an arbitrary field 
content.

There are many applications of the ideas described above.  For instance, 
we can apply it to product-group theories~\cite{Yanagida:1994vq,%
Izawa:1997he,Weiner:2001pv}.  Let us once again consider 
a supersymmetric $SU(5)$ gauge theory in the 5D warped spacetime 
of Eq.~(\ref{eq:metric}).  We then introduce an additional gauge 
group $SU(3) \times SU(2) \times U(1)$ on the IR brane, with the Higgs 
doublets charged under this IR-brane gauge group (and thus without 
being accompanied by any partner).  Now, we can consider that our 
low-energy $SU(3)_C \times SU(2)_L \times U(1)_Y$ gauge group is 
a diagonal subgroup of the bulk $SU(5)$ and the IR-brane $SU(3) \times 
SU(2) \times U(1)$.  (This breaking can be caused by the VEV of an 
appropriate IR-brane localized field).  Then, if the gauge couplings, 
$\tilde{g}_a$ of the original $SU(3)$, $SU(2)$ and $U(1)$ are large, 
$\tilde{g}_a \approx 4\pi$ ($a = 1,2,3$), the low-energy MSSM gauge 
couplings are effectively unified at the scale where $SU(5) \times 
SU(3) \times SU(2) \times U(1) \rightarrow SU(3)_C \times SU(2)_L 
\times U(1)_Y$ occurs $\sim k'$, since the gauge couplings of $SU(3)_C$, 
$SU(2)_L$ and $U(1)_Y$, $g_a$, are given by $1/g_a^2 = 1/g_5^2 + 
1/\tilde{g}_a^2 \approx 1/g_5^2$ at that scale.  Here, $g_5$ ($= O(1)$) 
is the coupling of the zero mode of the bulk $SU(5)$ gauge field. 
A problem of the corresponding scenario in 4D~\cite{Weiner:2001pv} 
is that, since the original $U(1)$ gauge coupling is strong at the 
unification scale, it hits the Landau pole immediately above that scale. 
There, it is also not clear why the three independent gauge couplings 
of $SU(3)$, $SU(2)$ and $U(1)$ become strong at a single scale, which 
must also coincide with the scale of diagonal breaking to avoid large 
threshold corrections.  Our theory addresses all of these issues 
naturally --- since the $SU(3)$, $SU(2)$ and $U(1)$ gauge fields are 
composite states of the GUT breaking dynamics, they all have strong 
couplings, $\tilde{g}_a \approx 4\pi$, at the scale where breaking 
to the diagonal subgroup occurs, and there is no issue of a Landau 
pole above this scale.  A potentially large mass for the Higgs 
doublets can be avoided by introducing a (discrete) gauge symmetry 
with the anomalies canceled by the Green-Schwarz terms on the IR brane.%
\footnote{Another possibility for canceling anomalies is to add extra 
matter fields (in complete $SU(5)$ multiplets) that obtain masses from 
supersymmetry breaking~\cite{Kurosawa:2001iq}.  We thank N.~Maru for 
pointing out this work to us.}
(To do all of these completely within the regime of effective field 
theory, the scale of diagonal breaking should be somewhat below the 
IR-brane cutoff, and the $SU(3)$, $SU(2)$ and $U(1)$ gauge couplings 
should be asymptotically non-free.  These can be arranged with an 
appropriate introduction of massive fields on the IR brane.  In the 
limit that the scale of diagonal breaking approaches the IR-brane 
cutoff, this theory is reduced to one of the theories discussed in 
the second paragraph of this subsection, where the Higgs doublets 
are located on the GUT-breaking IR brane.)  Note that since the quarks 
and leptons are introduced in the bulk in representations of the 
bulk $SU(5)$, we are still considering a unified theory of quarks 
and leptons (although it is possible to introduce them on the IR brane 
in representations of $SU(3) \times SU(2) \times U(1)$).  In particular, 
proton decay from unified gauge boson exchange still exists.  The Yukawa 
couplings of matter to the Higgs fields arise through the IR-brane 
VEV, breaking $SU(5) \times SU(3) \times SU(2) \times U(1)$ down to 
$SU(3)_C \times SU(2)_L \times U(1)_Y$. 

In most of the theories described above, the Higgs fields are localized 
to the IR brane, so that they are composite fields of the dynamical 
GUT breaking sector in the 4D description.  (This need not be the case. 
One of the theories described in the second paragraph of this subsection 
contains Higgs fields that propagate in the bulk, with appropriate 
boundary conditions imposed at the IR brane.  We can, however, always 
localize them to the IR brane by introducing appropriate hypermultiplet 
masses.)  The observed hierarchies of quark and lepton masses and mixings 
can then always be explained by wavefunction overlaps between the matter 
and Higgs fields, by appropriately choosing the bulk hypermultiplet 
masses for the matter fields such that lighter generations are localized 
more towards the UV brane, as e.g. in Eq.~(\ref{eq:c-flavor}).  (The 
option of localizing lighter generations towards the IR brane is not 
available unless the Higgs fields are pseudo-Goldstone boson multiplets.)
We find it very interesting that our framework of ``holographic grand 
unification'' accommodates many different ideas of solving the problems 
of conventional SUSY GUTs, developed mainly in the 4D context, with 
the automatic bonus of explaining the observed hierarchies of fermion 
masses and mixings through the wavefunction profiles of matter fields 
in the extra dimension.

\section{Discussion and Conclusions}
\label{sec:concl}

In this paper we have studied a framework in which grand unification 
is realized in truncated warped higher dimensional spacetime, 
where the UV and IR branes set the Planck and unification scales, 
respectively.  In the 4D description, this corresponds to theories 
in which the the grand unified gauge symmetry is spontaneously 
broken by strong gauge dynamics having a large 't~Hooft coupling, 
$\tilde{g}^2 \tilde{N}/16\pi^2 \gg 1$ (and a large number of ``colors'', 
$\tilde{N} \gg 1$).  In this parameter region, an appropriate (weakly 
coupled) description of physics is obtained in higher dimensions, and 
physics above the unification scale is determined by higher dimensional 
field theories, e.g. by specifying the spacetime metric, gauge group, 
matter content, boundary conditions, and Lagrangian parameters.  This 
allows us to control certain dynamical properties of the GUT breaking 
sector in the regime where effective field theory applies.  For example, 
we can make the size of threshold corrections small by making the 
symmetry breaking VEV on the IR brane (slightly) smaller than its naive 
value.  Moreover, the framework allows us to straightforwardly adopt 
intuitions and mechanisms arising from the higher dimensional picture. 
In particular, we can explain the observed hierarchies in quark and 
lepton masses and mixings in terms of the wavefunction profiles of 
matter fields in higher dimensions.  The generated hierarchies are 
naturally of the right size, of order $M_U/M_* \simeq 1/20$.

We have presented several realistic models within this framework.  In 
one model, on which we have focused the most, the lightness of the Higgs 
doublets is explained by the pseudo-Goldstone mechanism.  The strong 
gauge dynamics sector possesses a global $SU(6)$ symmetry as a ``flavor'' 
symmetry, of which the $SU(5)$ ($\times U(1)$) subgroup is gauged and 
identified as the unified gauge group.  When the global symmetry is 
broken dynamically to $SU(4) \times SU(2) \times U(1)$, the unified 
gauge symmetry is broken to the standard model gauge group, and the two 
MSSM Higgs doublets arise as massless pseudo-Goldstone supermultiplets. 
In our framework, this is realized by postulating a bulk $SU(6)$ gauge 
symmetry, broken to $SU(5) \times U(1)$ and $SU(4) \times SU(2) \times 
U(1)$ on the UV and IR branes, respectively.  One of the difficulties 
in implementing this mechanism in the conventional 4D framework is to 
find a way to suppress effects of explicit breaking in the potential 
generating the spontaneous $SU(6)$ breaking, since such effects would 
reintroduce an unacceptably large mass for the Higgs doublets.  In our 
case, these effects are (exponentially) suppressed by a large mass 
dimension for the operator generating the spontaneous $SU(6)$ breaking. 
Such an assumption is easy to implement in higher dimensions -- simply 
assume that the Higgs field breaking $SU(6)$ is localized to the IR 
brane.  This provides another example of the ``controllability'' of 
strong gauge dynamics in the large 't~Hooft coupling regime.

We have also demonstrated that many ideas for solving the problems of 
conventional 4D SUSY GUTs can be naturally implemented on the IR brane. 
We have presented several realistic models of this kind, for example, 
ones based on missing partner type or product group type scenarios. 
These models have the interesting feature that the GUT scale physics 
on the IR brane does not affect physics at higher energies, since the 
relevant physics arises as a result of the strong GUT breaking dynamics 
(as composite states) in the 4D description.  For example, large 
GUT multiplets, often needed to solve the problems of SUSY GUTs, 
do not contribute to the evolution of the unified gauge coupling 
at higher energies, and gauge fields having very large gauge couplings 
can naturally arise at the GUT scale without having the problem of 
a Landau pole.  These features open up new possibilities for GUT model 
building.

One can view the ``success'' of the present framework in several 
different ways.  For one who is interested in addressing the 
phenomenology of unified theories, such as gauge coupling unification 
and proton decay, models in our framework can be used to give predictions 
of observable quantities.  For example, we can explore relations 
between the branching ratios of proton decay and matter configurations 
in the extra dimension, as in the case of unified theories in flat 
space~\cite{Nomura:2001tn}.  Models of fermion masses and mixings, 
as well as models of supersymmetry breaking, can also be developed 
within the framework.%
\footnote{For example, we can take one of the supersymmetry breaking 
models in~\cite{Goldberger:2002pc}, with the boundary conditions at 
the UV brane changed to be trivial, and glue that spacetime (the 5D 
warped spacetime with the scales at the UV and IR branes taken to be the 
Planck and TeV scales, respectively) to one of our holographic warped 
GUT spacetimes discussed in section~\ref{sec:other}, at the UV branes 
of both spacetimes.  (The consistency of such constructions in effective 
field theory has been discussed recently in~\cite{Cacciapaglia:2006tg}.) 
In the 4D description, this corresponds to the situation where both 
the unified gauge symmetry and supersymmetry are broken by strong 
gauge dynamics, at the unification scale and the TeV scale, respectively. 
In practice, this system is analyzed most efficiently by first 
integrating out the GUT scale physics.  Then the low energy effective 
theory is simply reduced to one of the models in~\cite{Goldberger:2002pc}, 
but now we have an understanding of the hierarchical structure of the 
Yukawa couplings, located on the UV brane of the effective theory. 
While this effective field theory may be at the border of the weak 
and strong coupling regimes in 5D, it may still reproduce gross 
features of physical quantities, e.g. the superparticle spectrum, 
as is the case in higher dimensional formulations of QCD.}
On the other hand, one may be interested in exploring possible 
``UV completions'' of models formulated in warped spacetime.  It is 
possible, after all, that there may be some nontrivial consistency 
conditions in higher dimensional field theories, which are difficult 
(though not impossible) to catch in effective theory, and one way 
of ensuring the consistency of such theories is to ``derive'' them 
from complete UV theories.  Such ``UV completions'' may be achieved, 
for example, by embedding models into string theory, identifying 
a ``dual'' 4D theory, or by finding a 4D theory whose infrared 
fixed point has similar features as the original models in warped 
space~\cite{Arkani-Hamed:2001ca}.  From this perspective, our framework 
offers a guide on which models ``UV theorists'' should aim to reproduce; 
for example, string theorists may want to reproduce unified theories 
in 5D warped spacetime, with the unified gauge symmetry broken at 
an IR throat, rather than 4D unified theories directly from 
compactification.

We finally comment on the possibility that the unified gauge 
symmetry is broken by strong gauge dynamics whose 't~Hooft coupling 
is large but not extremely large, i.e. $\tilde{g}^2 \tilde{N}/16\pi^2 
\sim 1$.  In this case, the picture based on higher dimensional 
spacetime is not fully justified, but even then some properties of 
theories, especially properties associated with the IR brane physics 
(GUT breaking dynamics), may be effectively described by our higher 
dimensional warped unified theories.  In fact, such an approach had 
a certain level of successes in describing physics of lowest-lying 
excitations in QCD~\cite{Erlich:2005qh}.  In this sense, our framework 
may have a larger applicability than what is naively expected.

In summary, we have presented a framework in which dynamical GUT 
breaking models are realized in a regime that has a weakly coupled 
``dual'' picture.  Grand unified theories are realized in warped 
higher dimensional spacetime, with the UV and IR spacetime cutoffs 
providing the Planck and the unification scales, respectively. 
Several types of realistic models are discussed, with interesting 
implications for quark and lepton masses and mixings.  It would 
be interesting to study further implications of these models, such 
as those on proton decay, precise gauge coupling unification, 
supersymmetry breaking, and flavor physics.

\section*{Acknowledgments}

This work was supported in part by the Director, Office of Science, Office 
of High Energy and Nuclear Physics, of the US Department of Energy under 
Contract DE-AC02-05CH11231.  The work of Y.N. was also supported by the 
National Science Foundation under grant PHY-0403380, by a DOE Outstanding 
Junior Investigator award, and by an Alfred P. Sloan Research Fellowship.

\end{document}